# pgMAP: a pipeline to enable guide RNA read mapping from dual-targeting CRISPR screens

Phoebe C. R. Parrish[1,2,3], Daniel J. Groso[1,2], James D. Thomas[2,4], Robert K. Bradley[2,3,4], Alice H. Berger[1,2,3,5]

[1] Human Biology Division, Fred Hutchinson Cancer Center, Seattle, WA 98109, USA

[2] Computational Biology Program, Public Health Sciences Division, Fred Hutchinson Cancer Center, Seattle, WA 98109, USA

[3] Department of Genome Sciences, University of Washington, Seattle, WA 98195, USA

[4] Basic Sciences Division, Fred Hutchinson Cancer Center, Seattle, WA 98109, USA

[5] Lead Contact

**Corresponding author: Alice Berger (ahberger@fredhutch.org)**

**Abstract**

*Results*
We developed pgMAP, an analysis pipeline to map gRNA sequencing reads from dual-targeting CRISPR screens. pgMAP output includes a dual gRNA read counts table and quality control metrics including the proportion of correctly-paired reads and CRISPR library sequencing coverage across all time points and samples.

*Availability and Implementation*
pgMAP is implemented using Snakemake and is available open-source under the MIT license at https://github.com/fredhutch/pgmap_pipeline.

*Contact*
Dr. Alice Berger, ahberger@fredhutch.org

---

# 1. Introduction

Cancer therapies that target synthetic lethal interactions have the potential to expand treatment options for patients. In recent years, several genome-scale, dual-targeting CRISPR-mediated knockout (KO) screening approaches have been developed to map genetic interactions (GIs) in the human genome (Dede *et al.*, 2020; Gonatopoulos-Pournatzis *et al.*, 2020; Ito *et al.*, 2021; Köferle *et al.*, 2022; Parrish *et al.*, 2021; Tang *et al.*, 2022; Thompson *et al.*, 2021). These methods enable functional profiling of duplicated gene families and expand the range of potentially targetable synthetic lethal interactions in cancer (Dandage and Landry, 2021; Ryan *et al.*, 2023). However, computational methods for GI mapping from human CRISPR screen data are poorly established, which may impede interpretation of dual-targeting CRISPR screen data and thus prevent identification of actionable synthetic lethal targets.

Commonly-used tools for analyzing single-targeting CRISPR KO screen data, e.g. MAGeCK (Li *et al.*, 2014) and BAGEL (Hart and Moffat, 2016), have limited utility for multi-targeting CRISPR screens. These tools can be used to quantify the fitness effects of each dual gRNA construct once sequencing reads have been mapped to a reference. However, neither MAGeCK nor BAGEL can be used to directly map sequencing reads generated by dual-targeting CRISPR approaches, nor can they map GIs since they were designed to be used only with single-targeting CRISPR approaches.

Some software packages to analyze sequencing data from dual-targeting CRISPR screens have been developed (Ward *et al.*, 2021; Zamanighomi *et al.*, 2019). However, these methods have some drawbacks. One method requires a counts table as input, meaning users must use custom code to map their dual gRNA sequencing reads prior to using the tool (Zamanighomi *et al.*, 2019). Moreover, since both existing tools are implemented as R packages, their dependencies are not version-controlled and individual scripts must be run one-by-one, which limits scalability, reproducibility, and ease-of-use for users with limited computational experience.

Here, we present pgMAP (paired guide RNA mapper), a Snakemake-based analysis pipeline to map sequencing reads and generate a dual gRNA counts table from dual-targeting CRISPR-mediated KO screening data (**Figure 1A**). This read-mapping strategy was originally designed to analyze data from the pgPEN (paired guide RNAs for paralog genetic interaction mapping) screening method (Parrish *et al.*, 2021). pgMAP represents several improvements to the code base, including the use of the Snakmake workflow manager (Mölder *et al.*, 2021) and Conda environments (Anaconda Software Distribution, 2016) to ensure scalable and reproducible analysis. pgMAP also calculates and outputs summary statistics to enable quality control analysis. pgMAP will allow users to apply pgPEN to model systems of interest and enable discovery of genetic interactions across many organisms, genetic backgrounds, tissues, and cancer types.

## 2. Materials and methods

pgMAP computes dual gRNA counts and quality control (QC) statistics from reads generated by the pgPEN Illumina sequencing strategy in an automated fashion. Reads are trimmed with FastX-Toolkit v0.0.14 (http://hannonlab.cshl.edu/fastx_toolkit/) and demultiplexed using idemp (https://github.com/yhwu/idemp). Next, demultiplexed reads are mapped to the pgPEN library reference with Bowtie v1.2.2 (Langmead *et al.*, 2009), generating files of alignments in Sequence Alignment/MAP (SAM) format. The SAM files are sorted and converted to Binary Alignment/Map (BAM) format using SAMtools (Li *et al.*, 2009).

In the last steps of the pipeline, dual gRNAs present in each sample are counted and the percentage of correctly-paired dual gRNAs is computed using `counter_efficient.R`, an R script (https://www.R-project.org) that incorporates the libraries Tidyverse v1.2.1 (Wickham *et al.*, 2019) and Rsamtools v1.34.1 via Bioconductor (Huber *et al.*, 2015). The final step of the pipeline runs a Python script `combine_counts.py` to generate a file that contains dual gRNA counts for each time point, replicate, and condition in tab-separated values (TSV) format.

## 3. Usage and examples

To begin, clone the repository from https://github.com/FredHutch/pgMAP_pipeline into an empty folder on a high-performance computing (HPC) cluster. pgMAP is run with Snakemake and requires an up-to-date version of the Mamba package manager (https://github.com/mamba-org/mamba) and an associated Conda environment containing Snakemake and its dependencies. The Conda environment for running pgMAP will be automatically deployed by the Bash script `run_snakemake.sh`, and can be found via the following path: `workflow/envs/snakemake.yaml`. Additional instructions and links for installation can be found in the file `README.md`, located in the base folder or on the homepage of the GitHub website.

The config folder contains sample versions of the configuration files required to run the Snakemake pipeline, which can also be used to analyze the tutorial dataset included in the pgMAP package (tutorial data is located in the folder `input/tutorial-fastqs/`). Information on running the tutorial and analyzing sequencing data from pgPEN CRISPR screens can be found in the file `README.md`. For more information on the pgPEN sequencing method, see: Parrish *et al.*, 2021, Figure S1A.

pgMAP is run on an interactive HPC node via the Bash script `run_snakemake.sh`, which activates the Conda environment and executes the pipeline:

```
snakemake --snakefile "workflow/Snakefile"  \
  --use-conda --conda-prefix "~/tmp/" --conda-frontend mamba \
  -k -p --reason --jobs 50 --latency-wait 80
```

Upon successful execution of the pgMAP pipeline, results sub-directories containing the output from each Snakemake rule will be generated (see **Figure 1B**). Dual gRNA count tables will be found in the `results/pgRNA_counts/` directory. Reports containing information about pipeline runtime and graphs of the Snakemake rules that were run will be generated in the `<workflow/reports/>` directory.

Example QC output from the pgPEN HeLa screen is shown in **Figure 1C**. This output includes per-sample statistics about dual gRNA pairing rates, runtime, and the fold-coverage of the reference dual gRNA CRISPR library based on the number of correctly-paired reads and the input library size.

## 4. Conclusion

We developed pgMAP, an analysis pipeline for aligning and counting Illumina sequencing reads from dual-targeting CRISPR screens to produce a dual gRNA counts table and quality control metrics. The pgMAP pipeline is designed to be implemented by computational scientists as well as bench researchers who are familiar with Linux-based command line interfaces and/or HPC environments and the Python programming language. pgMAP enables researchers with all levels of computational skills to apply dual-targeting CRISPR screening approaches to their own model systems of interest, offering novel insights into human genetic interactions on a genome scale.

**Acknowledgements**

The authors thank April Lo, Mitchell Vollger, and Robin Aguilar for advice on developing and implementing Snakemake pipelines.

*Funding*

This work was funded with support from the Lung Cancer Research Foundation and an American Cancer Society Research Scholar Grant RSG-21-09-01-ET. P.C.R.P. was supported by NSF DGE-1762114. D.J.G was supported in part by the NIH/NCI (R37 CA252050). A.H.B. was supported in part by the NIH/NCI (R00 CA197762 and R37 CA252050); the Devereaux Outstanding Investigator Award from the Prevent Cancer Foundation; the Stephen H. Petersdorf Lung Cancer Research Award; the Seattle Translational Tumor Research program; and the Innovators Network Endowed Chair. R.K.B. was supported in part by the NIH/NHLBI (R01 HL128239 and R01 HL151651); NIH/NCI (R01 CA251138); Edward P. Evans Foundation; Blood Cancer Discoveries Grant program through the Leukemia & Lymphoma Society, Mark Foundation For Cancer Research, and Paul G. Allen Frontiers Group (8023-20); Department of Defense Breast Cancer Research Program (W81XWH-20-1-0596); and the McIlwain Family Endowed Chair in Data Science. R.K.B. is a Scholar of The Leukemia & Lymphoma Society (1344-18). Sequencing was performed by the Fred Hutch Genomics Shared Resource (supported by NIH/NCI Cancer Center Support Grant P30 CA015704). Computational studies were supported in part by FHCRC's Scientific Computing Infrastructure (ORIP S10 OD028685).

*Data availability*

Raw and processed CRISPR screen sequencing data for the pgPEN CRISPR screens (GEO: GSE178179) is publicly available via GEO. All other data analyzed in this paper is available from the lead contact upon request.

*Conflicts of interest*

The authors declare no competing interests.

**Main Figure**

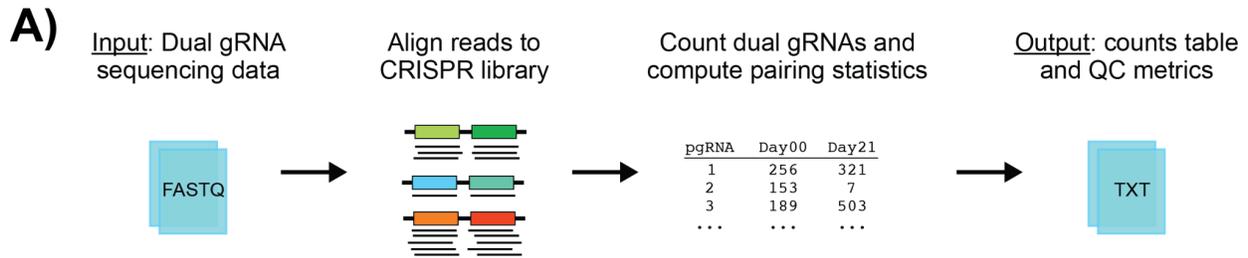

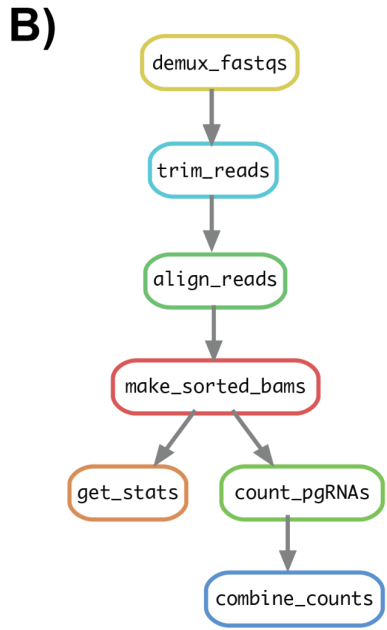

**Figure 1. pgMAP reproducibly and scalably maps sequencing reads from dual gRNA CRISPR screens.**

**(A)** Overview of pgMAP pipeline.

**(B)** Key rules run by Snakemake.

**(C)** Sample QC metrics for dual gRNA pairing and library coverage from the HeLa/iCas9 pgPEN CRISPR screen.

| Sample name | # correctly paired reads | % correctly paired reads | Library coverage | Runtime (mins.) |
|---|---|---|---|---|
| Plasmid | 38,471,205 | 78.84 | 1,160X | 120 |
| Day 5 (pooled) | 32,028,290 | 70.21 | 965X | 85 |
| Day 22, Rep A | 29,144,511 | 72.17 | 879X | 78 |
| Day 22, Rep B | 36,628,141 | 72.28 | 1,104X | 95 |
| Day 22, Rep C | 36,448,291 | 71.56 | 1,099X | 120 |